\documentclass[journal]{IEEEtran}
\usepackage{graphicx}
\usepackage{epstopdf}
\usepackage{multirow}
\usepackage{booktabs}
\usepackage{multicol, blindtext}
\usepackage{tabularx}
\usepackage{cite}
\usepackage{caption}
\usepackage{gensymb}
\usepackage{amsmath}
\usepackage{array}
\usepackage{float}
\usepackage{xcolor}
\usepackage{color, colortbl}
\pdfoutput=1
\usepackage[nottoc]{tocbibind}
 
\ifCLASSOPTIONcompsoc
\usepackage[caption=false,font=normalsize,labelfont=sf,textfont=sf]{subfig}
\else
  \usepackage[caption=false,font=footnotesize]{subfig}
\fi

\usepackage{dblfloatfix}
\usepackage[]{authblk}
\usepackage{url}

\author[1]{Sagar Dutta, \IEEEmembership{Student Member,~IEEE,}}
\author[1]{Banani Basu, \IEEEmembership{Senior Member,~IEEE,}}
\author[1]{Fazal A. Talukdar, \IEEEmembership{Member,~IEEE}}

\affil[1]{Department of Electronics and Communication Engineering \protect\\ National Institute Of Technology, Silchar, Assam, India}

\begin{document}

%content
%\tableofcontents
%\begin{figure*}
%\listoffigures
%\listoftables
%\end{figure*}

\title{Classification of Lower Limb Activities based on Discrete Wavelet Transform using On-Body Creeping Wave Propagation}
\maketitle

\begin{abstract}
This paper investigates how the creeping wave propagation around the human thigh could be used to monitor the leg movements. The propagation path around the human thigh gives information regarding leg motions that can be used for classification of activities. The variation of the transmission coefficient is measured between two on-body Polyethylene Terephthalate (PET) flexible antennas for six different leg based activities that exhibit unique time-varying signatures. A discrete wavelet transform (DWT) along with different classifiers like support vector machine (SVM), Decision Trees, Naive Bayes and K-nearest neighbours (KNN) is applied for feature extraction and classification to evaluate the efficiency for classifying different activity signals. Additional algorithms like dynamic time warping (DTW) and deep convolutional neural network (DCNN) has also been implemented, and in each case it SVM with DWT outperforms the others. Simulation to evaluate specific absorption rate (SAR) is carried out as the antenna is positioned on the human thigh leaving no gap. The results show that the SAR is within the threshold as per the FCC standard.
%A low power consumption and cost effective PET antennas with a dimension of 35.90 mm $\times$ 15.9 mm have been used as the sensing element for this study.
%due to its multi-resolution approach to evaluate the efficiency at classifying different activity signals. 
\end{abstract}


% For peer review papers, you can put extra information on the cover
% page as needed:
% \ifCLASSOPTIONpeerreview
% \begin{center} \bfseries EDICS Category: 3-BBND \end{center}
% \fi
%
% For peerreview papers, this IEEEtran command inserts a page break and
% creates the second title. It will be ignored for other modes.
\IEEEpeerreviewmaketitle

\section{Introduction}

In recent years, the advancement of wearable devices has empowered the development of wireless body area network (WBAN) and gained considerable interest in the electromagnetic (EM) wave propagation over the human body surface \cite{p1_a}. WBAN is a special network of sensors designed to link multiple wearable sensor nodes within and outside the human body autonomously. \textcolor{black}{There have been several studies carried out to investigate the use of complex body-wireless networks for the detection and recognition of human activities in remote health monitoring \cite{health1,health2}, activity tracking \cite{track1,track2}, security and surveillance \cite{sec1,sec2}, and human-computer interaction \cite{human1,human2}. In order to develop a robust and effective WBAN, it is important to understand how the EM waves propagate along or around the human body.} Researchers have carried out comprehensive studies over the last decade and have found that along-body propagation (line-of-sight), can be influenced by space wave and surface wave, while around-body propagation (non line-of-sight) is dominated by the creeping wave effects \cite{p1_c}.

Previous studies have focused primarily on the characteristics of wireless on-body propagation in order to classify human activities. Li et al. \cite{p2_a} utilized an on-body creeping wave around the human torso to classify compound activities such as hopping and sitting using a dynamic time warping algorithm and were able to achieve an average of 86\% accuracy. Bresnahan et al. \cite{p2_b} investigated the feasibility to monitor head and neck-based movements such as speaking, drinking, chewing, and deep breathing using EM creeping wave propagation around the human neck, by using deep convolutional neural network (DCNN) with a classification accuracy of over 80\%. Kim and Li \cite{p2_c,p2_c1} have used magnitude and phase of transmission coefficient ($S_{21}$) and reflection coefficient ($S_{11}$) for line-of-sight propagation between two on-body antennas to classify compound human activities. Xu et al. \cite{p2_d} classified finger movements based on reflection coefficient variation of a body-worn electrically small antenna and were able to achieve an accuracy of over 97\%. Moreover, Piuzzi et al. \cite{piuzzi2015complex} were able to achieve realistic scattering data of human subjects for the purpose of designing a UWB radar system for breath-activity monitoring. Pham et al. \cite{pham2017walking} proposed a walking monitoring system for the standard and front-wheel walkers, and accurately estimated four different walking styles. \textcolor{black}{Sardini et al. \cite{sardini2015wireless}} proposed a wireless instrumented crutch for gait monitoring to assess the contribution of upper limbs during walking. Trost et al. \cite{trost2014machine} proposed human activity classification based on wrist and hip worn accelerometer data.
%The above activity classifications were achieved by using antenna instead of sensors such as gyroscopes and accelerometers.
%The use of creeping wave propagation around the thigh to classify various leg movements has not been studied as per the author's knowledge.

\textcolor{black}{The objective of this paper is to investigate the feasibility of classifying various leg-based movements based on creeping wave propagation.} The leg's thigh region can reveal valuable physiological details about the motion of the leg and thus it has great potential for classification of leg-based movement. The proposed approach is to classify six different leg movements by using creeping wave propagation around the thigh. Both the magnitude and phase of the transmission coefficient ($S_{21}$) will be used for pattern recognition of the activities at 2.45 GHz using flexible on-body antennas. Furthermore, a discrete wavelet transform is applied to the signal to extract the approximation coefficients and detail coefficients to be used as features. The main advantage of DWT is that it provides a good time resolution at high frequencies and a good frequency resolution at low frequencies. Due to its high time-frequency localization capability, the DWT can reveal the local characteristics of the signal. This makes it particularly suitable for non-stationary signal analysis. Then we apply machine learning-based classifiers to classify different leg-based activities. We have used a one-vs-one strategy that divides the multi-class classification into one binary classification problem for each pair of classes. The results indicate that the creeping wave propagation around the thigh can effectively classify different activities with high accuracy with the help of DWT.

\section{Creeping Waves Around The Thigh}
Ryckaert et al. \cite{ryckaert2004channel} proposed the first on-body propagation of a creeping wave based on the finite-difference time-domain simulations. They observed that as the EM wave travels around the human torso, its magnitude decays exponentially. Alves et al. \cite{alves2010analytical} derived a simple analytical path loss model from the creeping wave theory, describing the path loss around a lossy dielectric cylinder, where the decaying factor for the path loss model is expressed as,
\begin{equation}
\alpha _{dB/cm}\propto \frac{f^{1/3}}{r^{2/3}}
\end{equation}
where \textcolor{black}{$r$} is the circumference of the body in cm and \textcolor{black}{$f$} is the operating frequency in MHz. It is observed that the decaying factor is inversely proportional to the circumference of the body.

To illustrate how the creeping wave propagates around the human thigh, the propagation characteristics around the thigh is measured when the subject is not performing any activity. \textcolor{black}{This is the first creeping wave propagation study around-thigh that is expected to differ significantly from previously reported around-neck and around-torso propagation as the muscle, fat, and skin tissue vary between the thigh and neck/torso. Moreover, flexible PET antennas having dimension 35.90 mm $\times$ 15.9 mm and operating at 2.45 GHz have been used for both transmitting and receiving antennas instead of monopole antennas used in previous studies \cite{p2_a,p2_b}.}

\begin{figure}[t]
\centering
\subfloat[]{\includegraphics[scale=0.03]{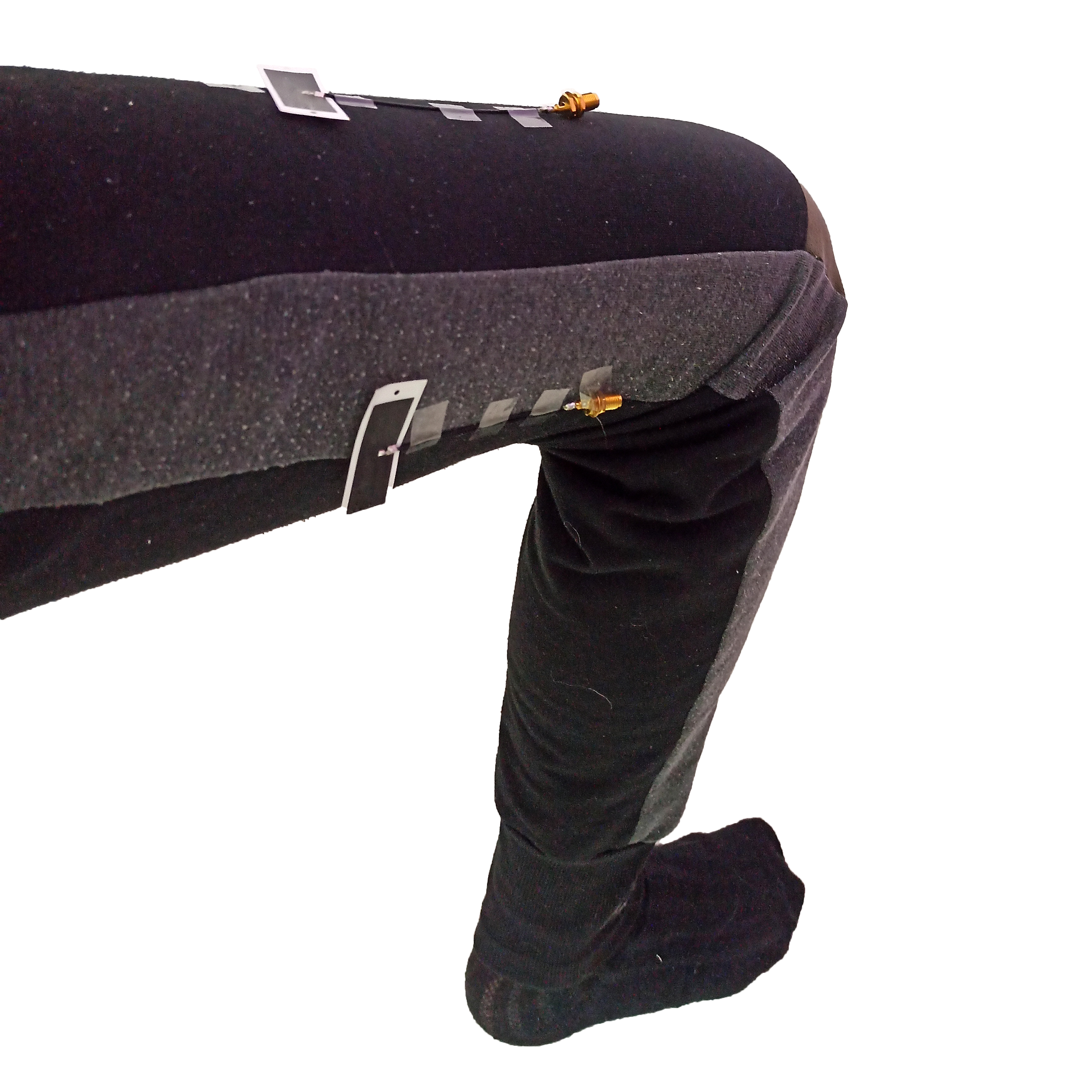}%
\label{f1a}}
\hfill
\subfloat[]{\includegraphics[width=1.7in]{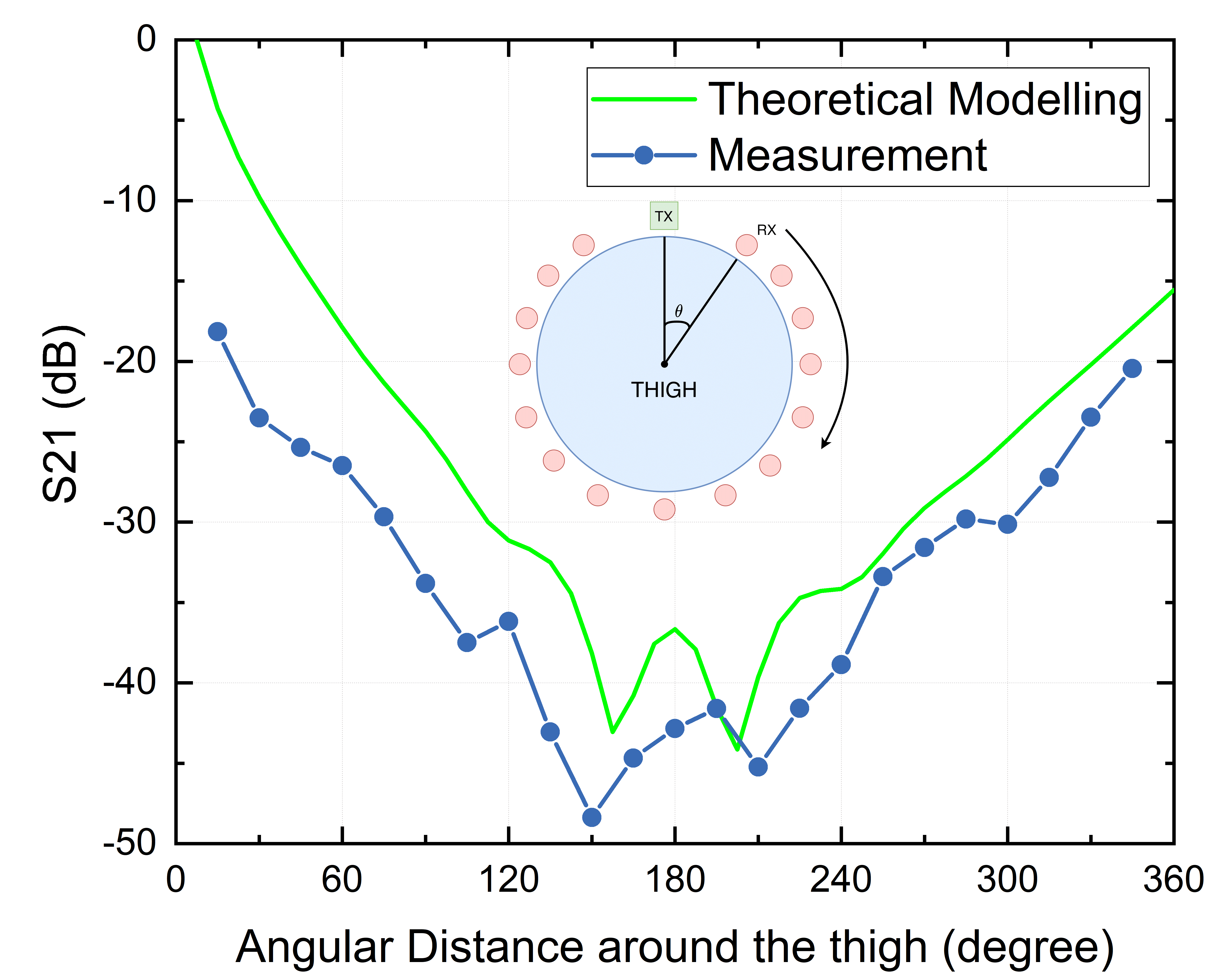}%
\label{f1b}}
\caption{(a) Measurement Setup (b) \textcolor{black}{Transmission loss around the thigh}}
\label{setup}
\end{figure}

The experimental setup of our investigation is shown in Fig. \ref{f1a}. \textcolor{black}{The power received at antenna 2 (Rx) relative to the power input to antenna 1 (Tx) is referred as transmission coefficient ($S_{21}$).} To measure the complex transmission coefficient data ($S_{21}$), a circular path around the thigh is selected.
%A circular path around the thigh is selected to measure the complex transmission coefficient data ($S_{21}$). 
The transmitting antenna remains fixed on the front of the right thigh and the receiving antenna moves around the thigh in a circular path with a \textcolor{black}{step size of 7.5\degree} (1 cm) keeping the transmitter and the receiver in the same horizontal plane. The human subject stands in parade rest position during the measurement to avoid any interference. Both antennas are connected to a two-port Anritsu MS2037C Network Analyzer with Anritsu 15NNF50-1.5B test port extension cable. The measurements have been carried out in an anechoic chamber to avoid any external radio interference.

The transmission coefficient ($S_{21}$) data measured around the thigh having  circumference of 48 cm, at each step is shown in Fig. \ref{f1b}. \textcolor{black}{The x-axis denotes the angular distance of the receiver from the fixed transmitter in a counter-clockwise direction along a circular path around the thigh.} It is observed from Fig. \ref{f1b} that the magnitude of transmission coefficient ($S_{21}$) falls exponentially from \textcolor{black}{14{\degree} (2 cm) to 150{\degree} (20 cm)}. However, constructive interference of the signal is seen at an angular distance from \textcolor{black}{150{\degree} (20 cm) to 195{\degree} (26 cm)} at the back of the thigh due to the creeping waves arriving from both counter-clockwise and clockwise direction around the thigh, and it rises exponentially from \textcolor{black}{210{\degree} (28 cm) to 345{\degree} (46 cm)}. The path gain expression \cite{alves2010analytical} shown in the plot (Fig. \ref{f1b}) manifests a good agreement between our measurement and the theoretical results.
\begin{figure}[t]
    \centering
    \includegraphics[width=3.5 in]{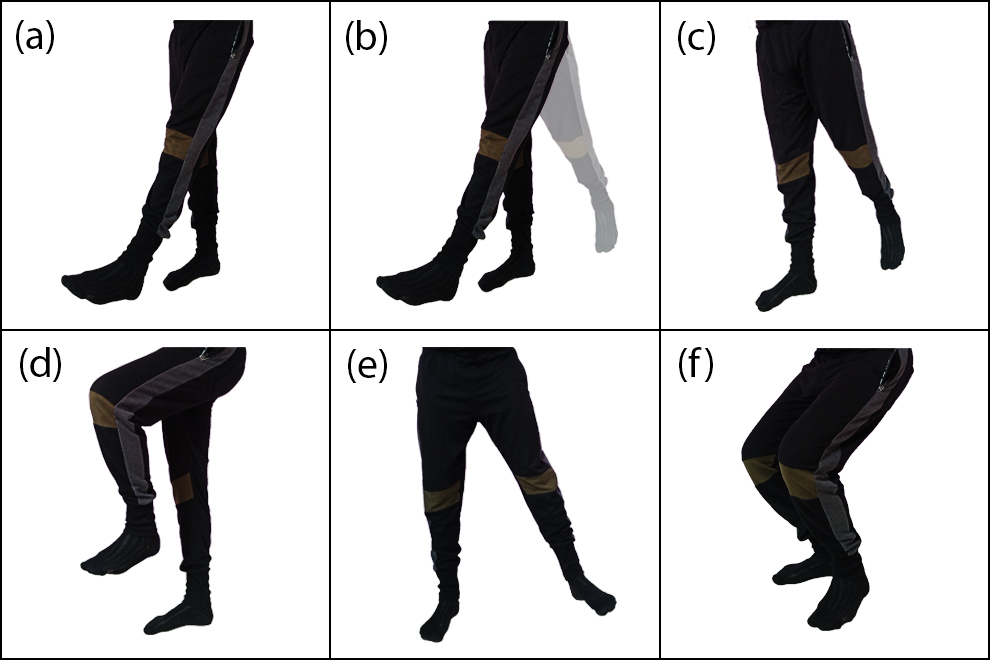}
    \caption{Six leg-based activities: (a) forward leg swing (b) full leg swing (c) backward swing (d) lifting knee (e) sideways leg swing (f) squatting}
    \label{act}
\end{figure}

%\begin{figure}[t]
%\centering
%
%\subfloat[]{\includegraphics[width=1.1in]{insert}%
%\label{a1}}
%\hfill
%\subfloat[]{\includegraphics[width=1.1in]{insert}%
%\label{a2}}
%\hfill
%\subfloat[]{\includegraphics[width=1.1in]{insert}%
%\label{a3}}
%\\
%\subfloat[]{\includegraphics[width=1.1in]{insert}%
%\label{a4}}
%\hfill
%\subfloat[]{\includegraphics[width=1.1in]{insert}%
%\label{a5}}
%\hfill
%\subfloat[]{\includegraphics[width=1.1in]{insert}%
%\label{6}}
%
%\caption{Six Leg-based activities (a) (b) (c) (d) (e) (f)}
%\label{act}
%\end{figure}

%\begin{figure*}[h!]
%    \centering
%    \includegraphics[width=6.5in]{activity2}
%    \caption{Measured transmission coefficient $S_{21}$ around thigh for six different activities at 2.45 GHz: (a) Forward leg swing (b) Complete leg swing (c) Backward leg swing (d) Lifting Knee (e) Sideways leg swing (f) Squatting}
%    \label{f3}
%\end{figure*}

%\begin{figure*}[t]
%    \centering
%    \subfloat{\includegraphics[width=3.5in]{male_act}%
%	}
%	\hfill
%	\subfloat{\includegraphics[width=3.5in]{female_act}%
%	}
%    \caption{Six Leg-based activities: (a) forward leg swing (b) full leg swing (c) backward swing (d) lifting knee (e) sideways leg swing (f) squatting}
%    \label{act}
%\end{figure*}
\begin{figure*}
    \centering
    \begin{minipage}{\columnwidth}
        \centering
        \includegraphics[width=0.9\textwidth]{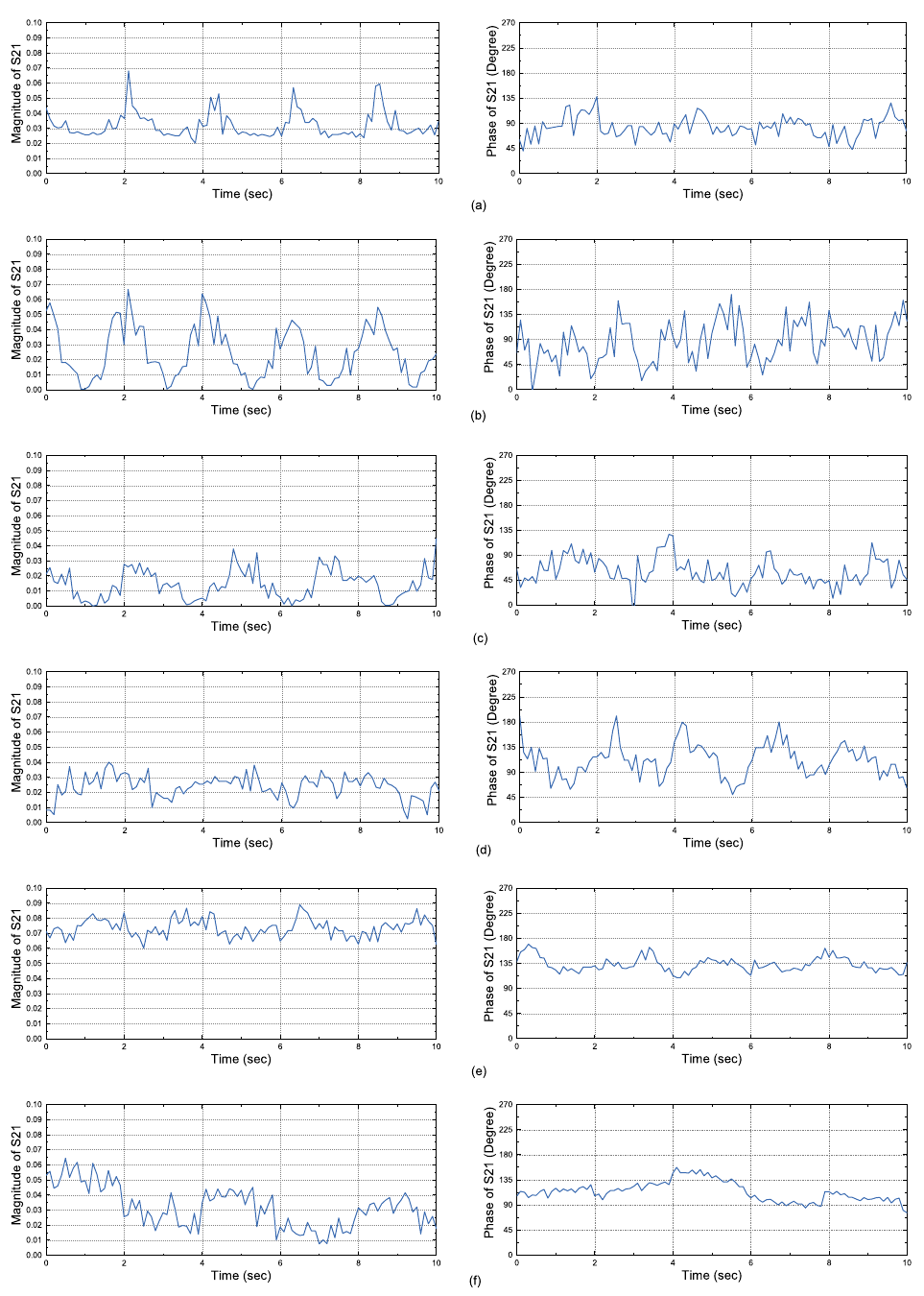} % first figure itself
        \caption{$S_{21}$ for six leg-based activities for representative male participant: (a) forward leg swing (b) full leg swing (c) backward swing (d) lifting knee (e) sideways leg swing (f) squatting}
        \label{f3}
    \end{minipage}\hfill
    \begin{minipage}{\columnwidth}
        \centering
        \includegraphics[width=0.9\textwidth]{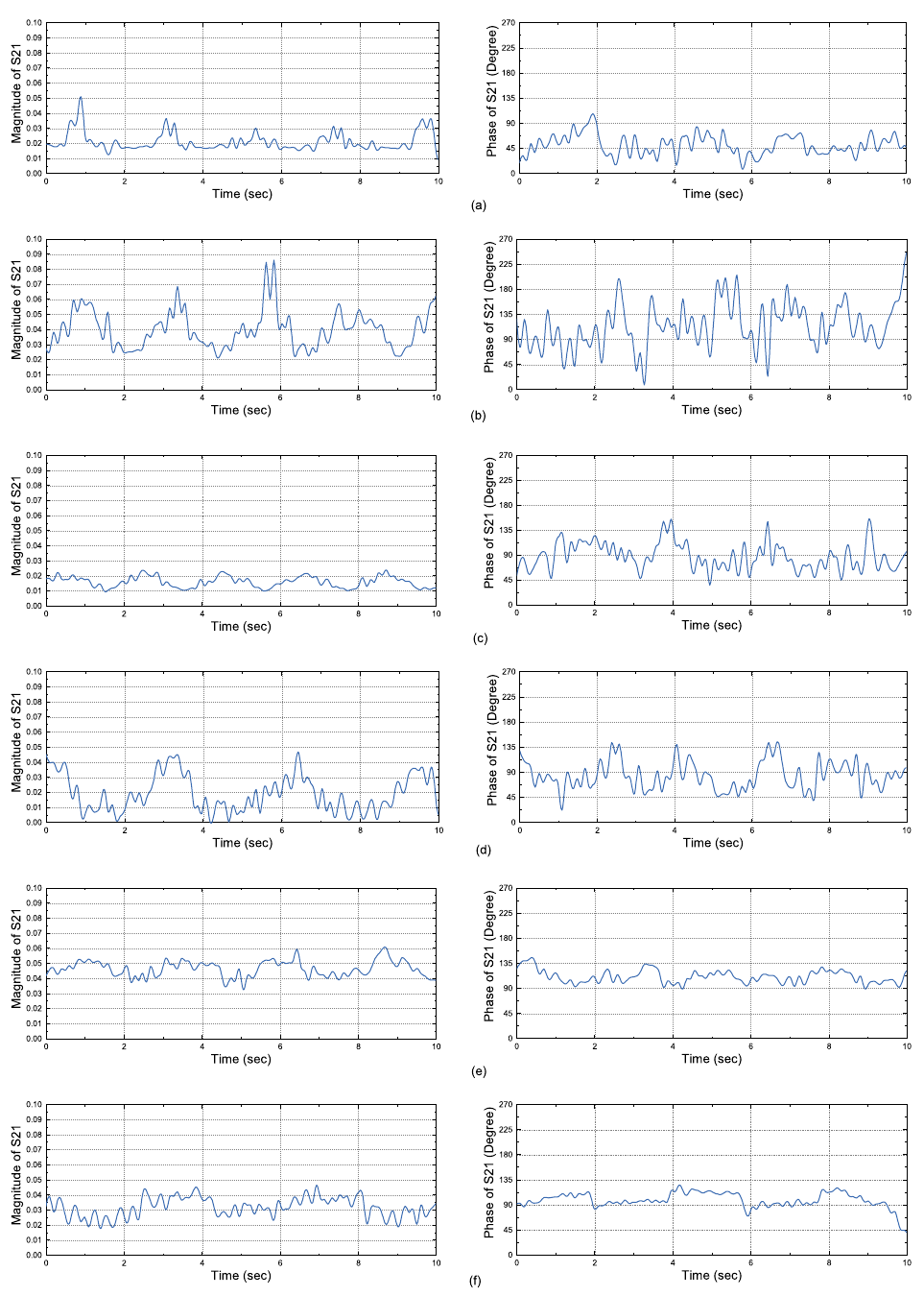} % second figure itself
        \caption{$S_{21}$ for six leg-based activities for representative female participants: (a) forward leg swing (b) full leg swing (c) backward swing (d) lifting knee (e) sideways leg swing (f) squatting}
        \label{f4}
    \end{minipage}
\end{figure*}
\section{Creeping wave measurement of leg based motions}
The measurement setup is similar to the setup in the previous section with the exception that \textcolor{black}{both the transmitter and receiver is kept fixed on the right thigh and the receiver is placed at an angular separation of $110\degree$ from the transmitter which is placed on the front side of the thigh as shown in Fig. \ref{f1a}.} \textcolor{black}{The placement of the antennas at different positions along the thigh is tried and the best result is found at the middle position of the thigh.} The same pair of PET on-body antennas operating at 2.45 GHz is used to cover the WBAN frequency band. Complex transmission coefficient data ($S_{21}$) are recorded for 20 seconds using a vector network analyzer (VNA) under continuous-time mode as the activities are being performed. The measurements were performed in an anechoic chamber. 

\textcolor{black}{Six participants were selected to perform six different right leg-based activities for the experiment. The physical specifications of the six participants are listed in Table \ref{tab3}.} The activities include forward leg swing, full leg swing, backward leg swing, lifting the knee, sideways leg swing, and squatting as illustrated in Fig. \ref{act}.

\begin{table}[]
\centering
\caption{Participant Specifications}
\resizebox{\columnwidth}{!}{%
\begin{tabular}{@{}lcccc@{}}
\toprule
& Height (cm) & Weight (kg) & Circumference of Thigh (cm) & Age (years) \\ \midrule
Male 1		  & 172         & 62          & 48                          & 27 \\ 
Male 2 		  & 175         & 80          & 54                          & 27 \\
Male 3 	 	  & 167         & 59          & 50                          & 25 \\
Female 1 	  & 162         & 54          & 45                          & 23 \\
Female 2 	  & 160         & 49          & 41                          & 25 \\ 
Female 3 	  & 157         & 43          & 43                          & 26 \\ \bottomrule
\end{tabular}
}
\label{tab3}
\end{table}

The propagation around the thigh is mainly influenced by the movement of the surrounding muscles and tissues which causes disturbances to form unique patterns in the measured $S_{21}$, which is used for the classification of different activities. Fig. \ref{f3} and Fig. \ref{f4} shows the time domain examples of the measured $S_{21}$ for the six different activities at 2.45 GHz for male and female participants respectively. The activities are performed in a periodic manner. Both the magnitude and phase data are observed to display unique patterns depending on activity, and can thus potentially be used to classify leg movements.
\section{Feature extraction and classification technique}
The feature extraction and classification are the two different stages involved in the proposed classification method to classify leg movements. Feature extraction technique based on discrete wavelet transform is applied to the signal to extract the wavelet coefficients to be used as features. The main advantage of DWT is that it provides a good time resolution at high frequencies and a good frequency resolution at low frequencies. Due to its high time-frequency localization capability, the DWT can reveal the local characteristics of the signal. This makes it particularly suitable for non-stationary signal analysis.

\begin{figure}[h!]
    \centering
    \includegraphics[width=2.6 in]{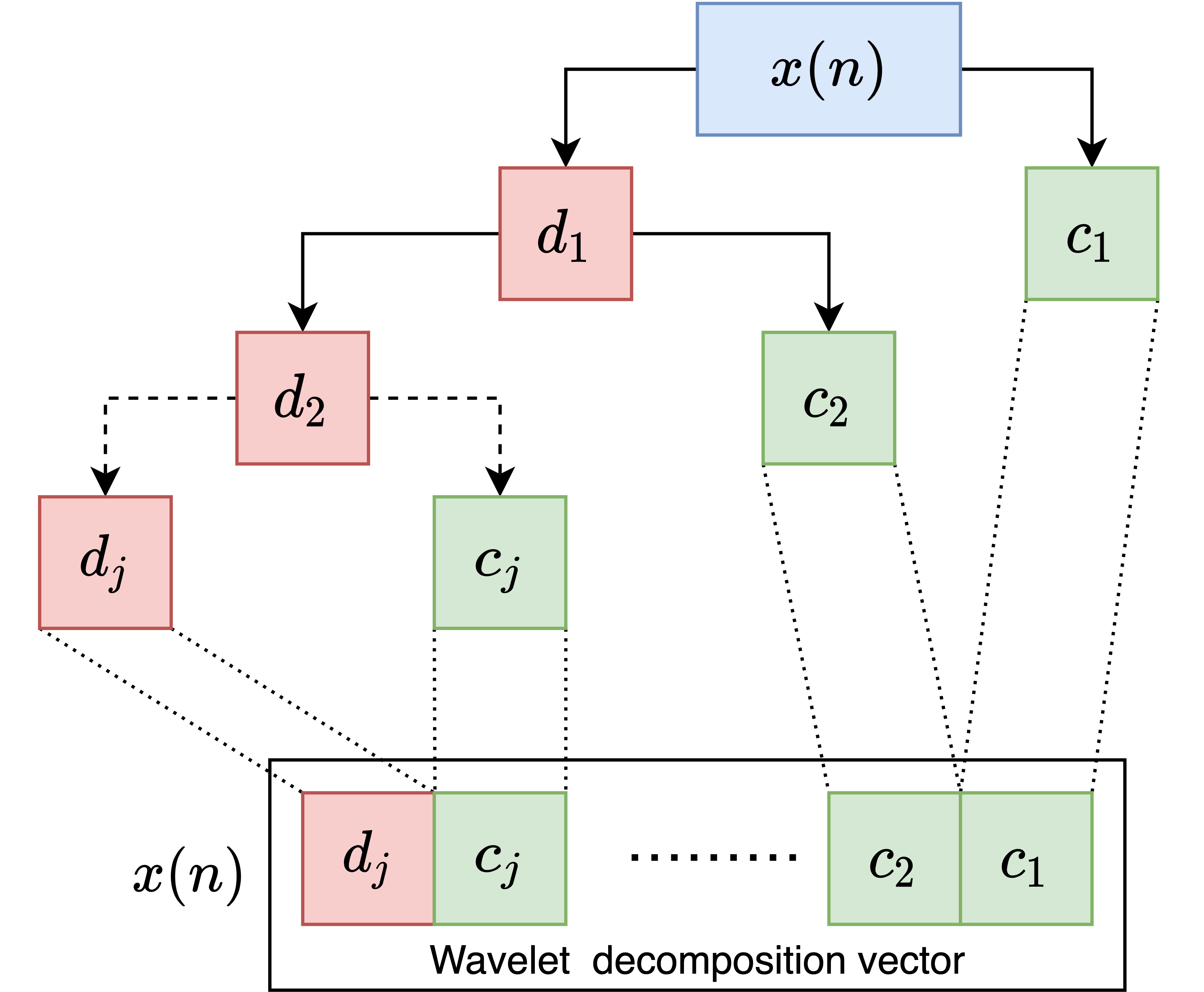}
    \caption{DWT decomposition upto j level}
    \label{f1}
\end{figure}

{\color{black}The wavelet transform (WT) transforms a signal into several wavelet basis functions using time-frequency decomposition. The continuous wavelet transform of a signal $s(t)$ is expressed as
\begin{equation}
WT(x,y)= \int_{-\infty}^{\infty}\psi_{(x,y)}^{*}(t)s(t)dt
\label{eq1}
\end{equation}
where * represents the complex conjugate and
\begin{equation}
\psi_{(x,y)}(t)= \frac{1}{\sqrt{x}}\psi \left ( \frac{t-y}{x} \right )
\label{eq2}
\end{equation}
is a member of mother wavelet function $\psi(t)$; scale and translation parameters are defined by 'x' and 'y' respectively. The DWT is derived by discretizing the wavelet $\psi_{(x,y)}(t)$ where 'x' and 'y' are replaced by $2j$ and $2^{j}k$ respectively and can be expressed as \cite{mallat1989theory}

\begin{equation}
D(j,k)= \int_{-\infty}^{\infty}\psi_{(j,k)}^{*}(t)s(t)dt
\label{eq3}
\end{equation}
where $\psi_{(j,k)}(t)= \frac{1}{\sqrt{2^{j}}}\psi  \left ( \frac{t-2^{j}k}{2^{j}} \right)$. The Mallat \cite{mallat1989theory} algorithm uses pyramidal structure to implement DWT, for each stage of wavelet decomposition, time dilation is carried out by downsampling. Wavelet filters are used for decomposition and reconstruction, and the scaling coefficients $d_{j}$ and the wavelet coefficients $c_{j}$ can be obtained by
\begin{equation}
d_{j}[s(t)]= \sum_{k}L(2t-k)d_{j-1}[s(t)]
\end{equation}

\begin{equation}
c_{j}[s(t)]= \sum_{k}H(2t-k)d_{j-1}[s(t)]
\end{equation}

%%%%%%%%%%%%%%%%%%%%%%%%%%%%%%%%%%%%%%%%%%%%%%%%%%%
\begin{figure*}[t]

\centering
\subfloat[]{\includegraphics[width=2.1 in]{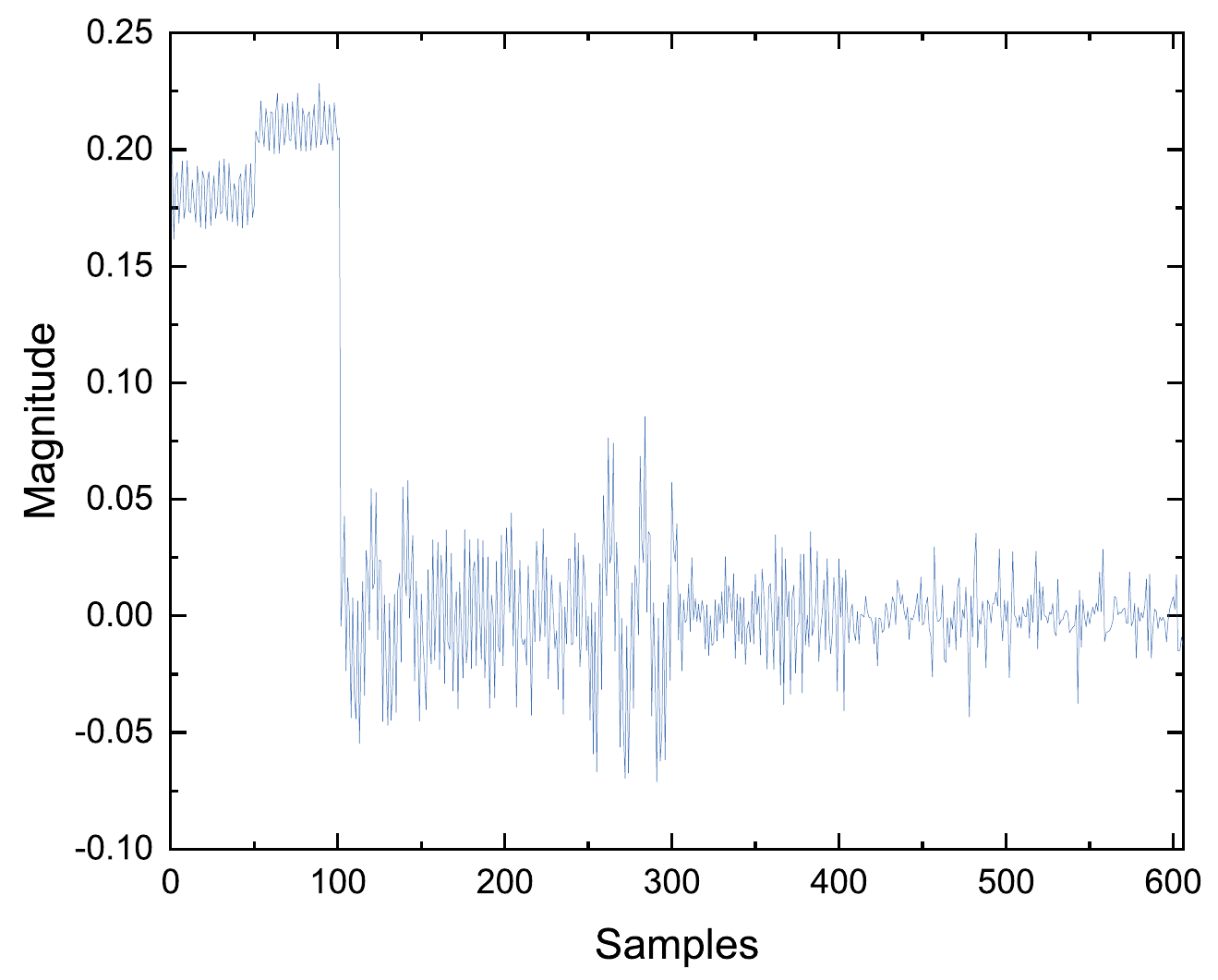}%
\label{d1}}
\hfill
\subfloat[]{\includegraphics[width=2.1 in]{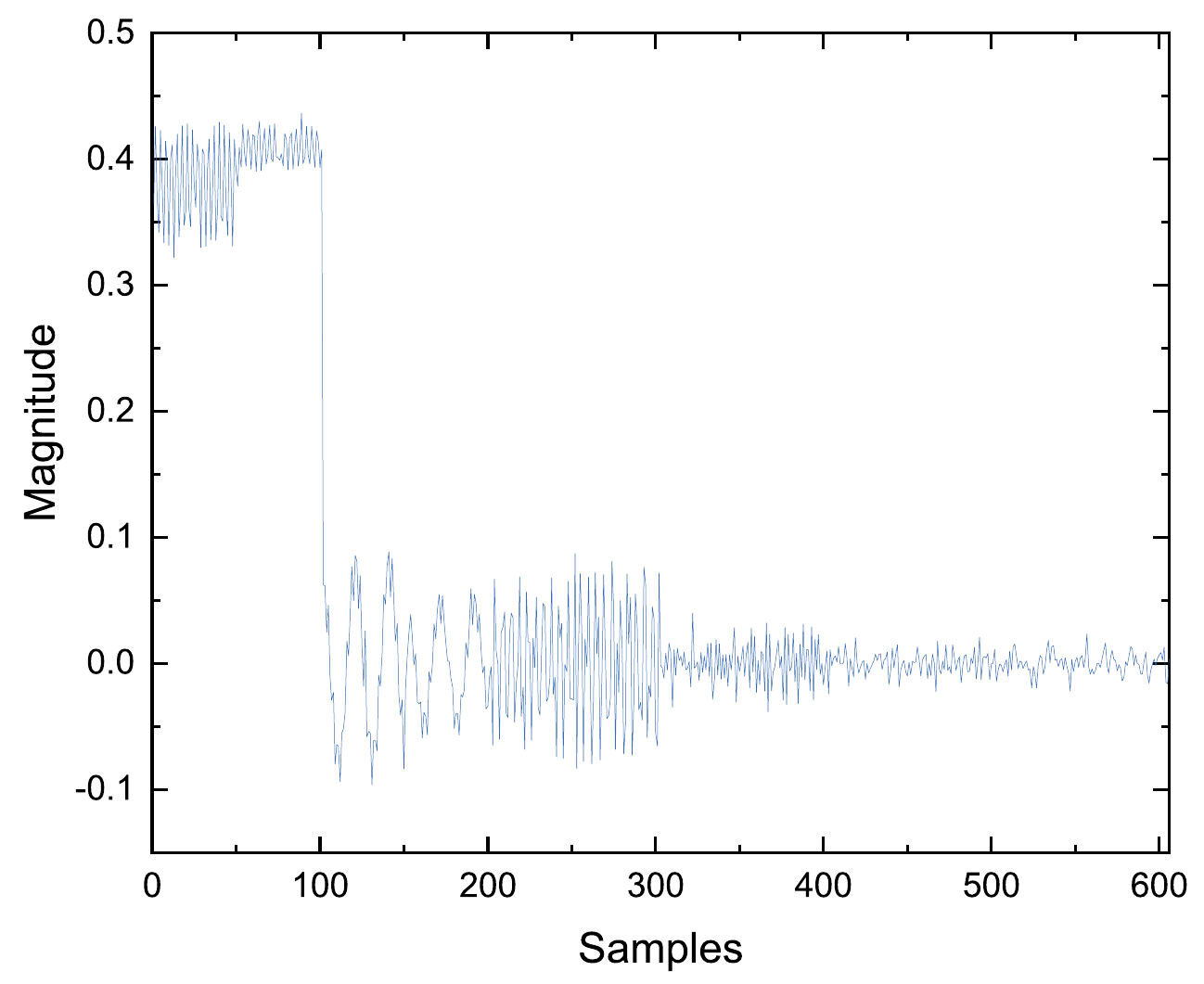}%
\label{d2}}
\hfill
\subfloat[]{\includegraphics[width=2.1 in]{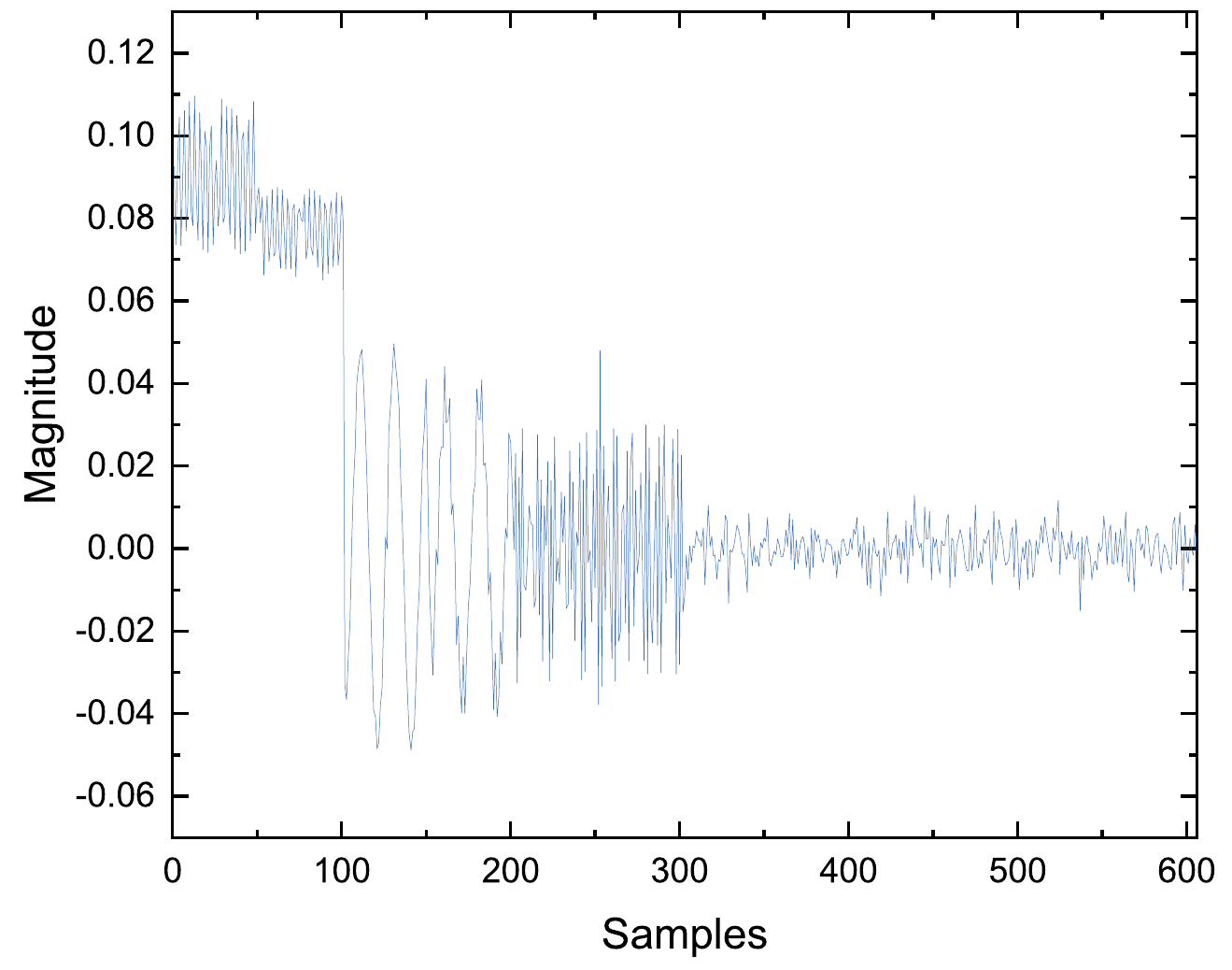}%
\label{d3}}
\\
\subfloat[]{\includegraphics[width=2.1 in]{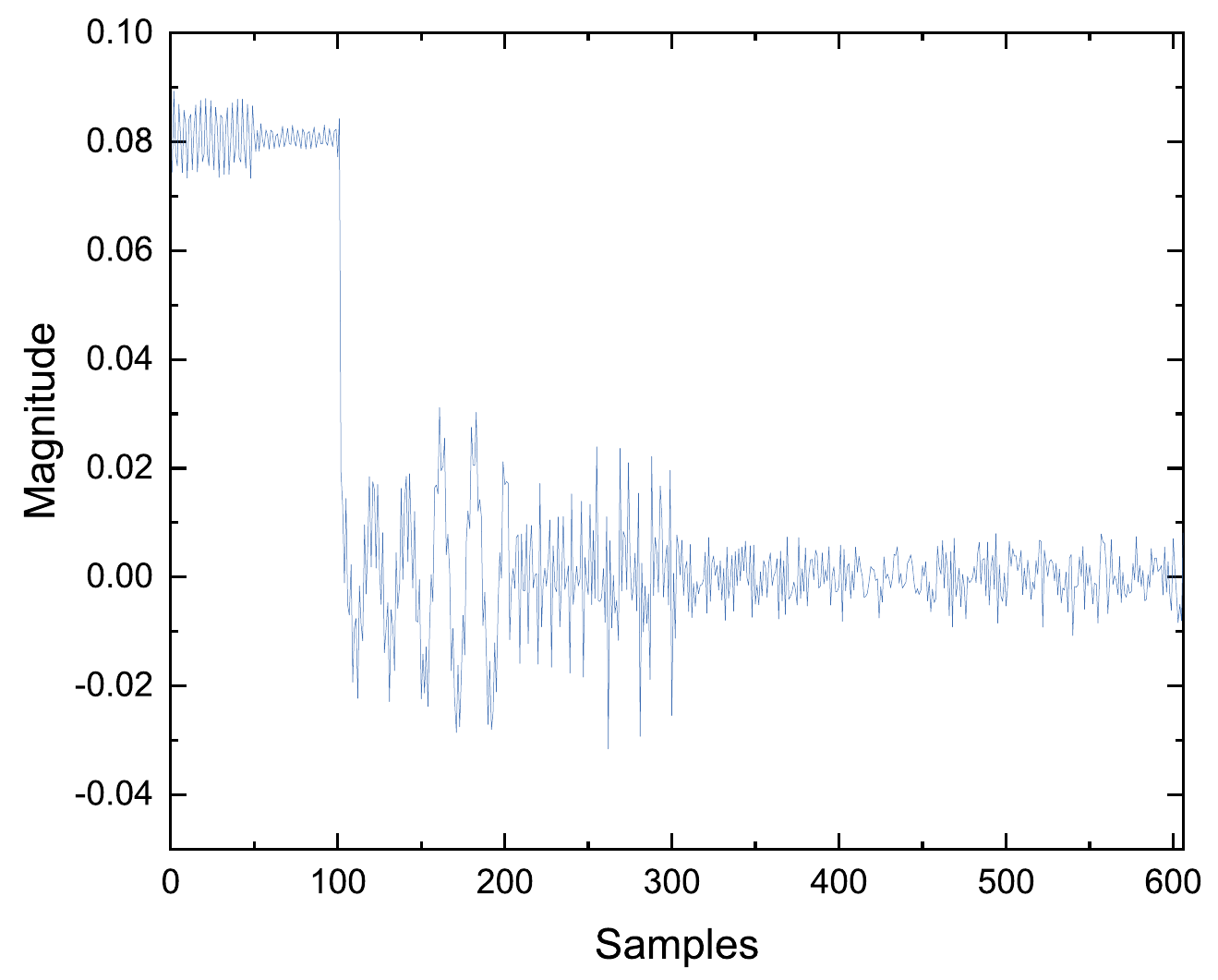}%
\label{d4}}
\hfill
\subfloat[]{\includegraphics[width=2.1 in]{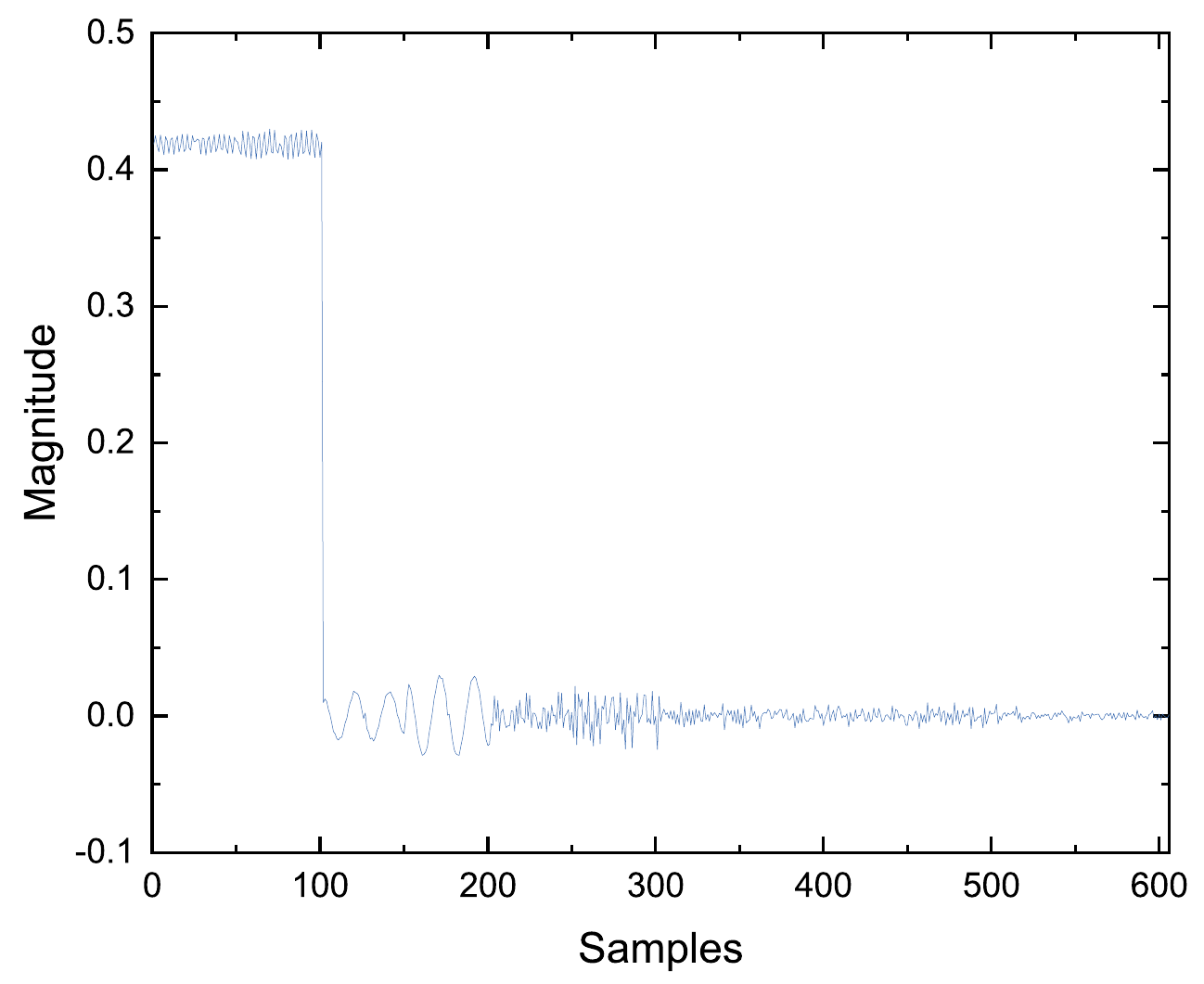}%
\label{d5}}
\hfill
\subfloat[]{\includegraphics[width=2.1 in]{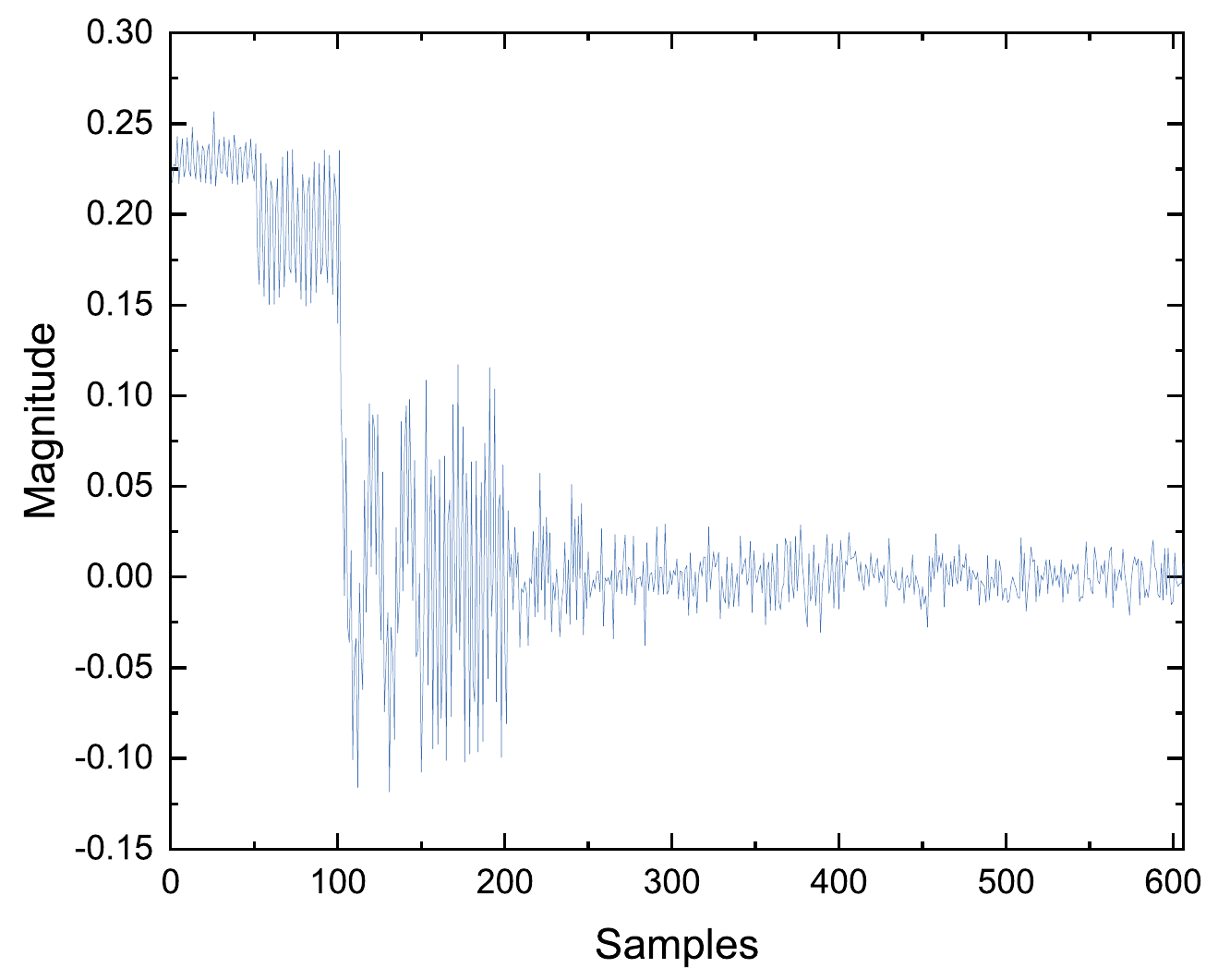}%
\label{d6}}
\caption{DWT coefficients of $S_{21}$: (a) Forward leg swing (b) Full leg swing (c) Backward leg swing (d) Lifting knee (e) Sideways swing (f) Squatting}
\label{dd}

\end{figure*}

%%%%%%%%%%%%%%%%%%%%%%%%%%%%%%%%%%%%%%%%%%%%%%%

\begin{table*}[!h]
\centering
\caption{\textcolor{black}{Training parameters}}
\resizebox{\textwidth}{!}{%
\begin{tabular}{|l|l|l|l|l|l|l|}
\hline
\multicolumn{1}{|c|}{Classifiers} & \multicolumn{1}{c|}{SVM}                                                                                             & \multicolumn{1}{c|}{KNN}                                                                            & \multicolumn{1}{c|}{Naive Bayes} & \multicolumn{1}{c|}{Decision Tree}                                                                     & \multicolumn{1}{c|}{DTW}                                                                                                                                                                                                                                                                                         & \multicolumn{1}{c|}{DCNN}                                                                                                                                                                                                                                                                                                                                                                                 \\ \hline
Parameters                        & \begin{tabular}[c]{@{}l@{}}Kernel function: Cubic\\ Box constraint level: 1\\ Multiclass method: 1 vs 1\end{tabular} & \begin{tabular}[c]{@{}l@{}}K: 10\\ Distance metric: Euclidean\\ Distance weight: Equal\end{tabular} & Distribution: Gaussian           & \begin{tabular}[c]{@{}l@{}}Number of Splits: 20\\ Split criterion: Gini's diversity index\end{tabular} & \begin{tabular}[c]{@{}l@{}}Cost function:\\ \\ for a= 1 to n\\    for b= 1 to m\\ \\ DTW(a,b)= d(x{[}a{]}, y{[}b{]}) + \\ min\{DTW(a-1, b), DTW(a, b-1), DTW(a-1, b-1)\}\\ \\ where DTW(a, b) = distance between signal \\ x{[}1:a{]} and y{[}1:b{]} and d(x{[}a{]},y{[}b{]}) = $\mid$x{[}a{]}-y{[}b{]}$\mid$\end{tabular} & \begin{tabular}[c]{@{}l@{}}Solver: SGDM\\ Batch size: 15\\ Initial learning rate: 0.001\\ Max epochs: 35\\ Activation function: ReLU\\ No. of hidden layers: 4\\ No. of fully connected layers: 2\\ Convolutional filter size: 3x3\\ Pooling layer reduction ratio: 2:1\\ Dropout rate: 0.5\\ Input: Spectrogram of raw signal\\ Input size: 100x100\\Training/Validation ratio: 70/30 \end{tabular} \\ \hline
\end{tabular}
}
\label{para}
\end{table*}

%%%%%%%%%%%%%%%%%%%%%%%%%%%%%%%%%%%%%%%%%%%%%%%%%%%%%%%

\begin{table*}[t!]
\centering
\caption{Mean value of the features for DWT coefficients for 4 sec time window}
%\resizebox{\columnwidth}{!}{%
\begin{tabular}{@{}lcccccc@{}}
\toprule
               & \begin{tabular}[c]{@{}c@{}}Mean\\ (Feature 1)\end{tabular} & \begin{tabular}[c]{@{}c@{}}Peak Value\\ (Feature 2)\end{tabular} & \begin{tabular}[c]{@{}c@{}}RMS Value\\ (Feature 3)\end{tabular} & \begin{tabular}[c]{@{}c@{}}Standard Deviation\\ (Feature 4)\end{tabular} & \begin{tabular}[c]{@{}c@{}}Kurtosis\\ (Feature 5)\end{tabular} & \multicolumn{1}{c}{\begin{tabular}[c]{@{}c@{}}Skewness\\ (Feature 6)\end{tabular}} \\ \midrule
Forward Swing  & 0.04                                                       & 0.26                                                             & 0.07                                                            & 0.08                                                                     & 3.72                                                           & 1.14                                                                               \\
Full Swing     & 0.06                                                       & 0.34                                                             & 0.18                                                            & 0.15                                                                     & 4.21                                                           & 1.50                                                                               \\
Backward Swing & 0.02                                                       & 0.08                                                             & 0.03                                                            & 0.05                                                                     & 3.35                                                           & 1.17                                                                              \\
Lifting Knee   & 0.01                                                       & 0.08                                                             & 0.05                                                            & 0.04                                                                     & 3.68                                                           & 1.25                                                                               \\
Sideways Swing & 0.08                                                       & 0.46                                                             & 0.13                                                            & 0.20                                                                     & 4.15                                                           & 1.58                                                                               \\
Squatting      & 0.02                                                       & 0.21                                                             & 0.09                                                            & 0.08                                                                     & 4.45                                                           & 1.84                                                                               \\ \bottomrule
\end{tabular}
%}
\label{tab1}
\end{table*}
where \textcolor{black}{$j$} is the decomposition level, \textcolor{black}{$L$} and \textcolor{black}{$H$} sequences are the low and high pass filters for wavelet decomposition resulting from the original mother wavelet $\psi(t)$. The decomposition of j-level DWT is shown in Fig. \ref{f1}.}

Different mother wavelets have different DWT coefficients for the same signal, which contribute to different detection performance. Discrete Meyer wavelet appears to produce the best filtering realization and has been used in this work with decomposition levels taken as five. For activity classification, it is important to choose appropriate features that best reflect the signal characteristics. The feature vector of each activity signal segment is composed of DWT coefficient features from many frequency bands. The feature vectors of each signal are built from DWT coefficients in multiple frequency bands. Examples of DWT coefficients of the measured S21 for the six activities are shown in Fig. \ref{dd}. The DWT coefficients have been extracted using the wavelet analyzer toolbox in MATLAB. However, we note that the distributions of the feature for the six activities are considerably overlapped with each other by analyzing the histograms (not shown here), and a more advanced classifier such as support vector machine (SVM) is needed.
\begin{table*}[t!]
\centering
\caption{\textcolor{black}{Performance comparison for 4 sec time window}}
\resizebox{\columnwidth}{!}{%
\begin{tabular}{@{}ccccc@{}}
\toprule
Classifier    & Accuracy & Precision & Recall & F1 score \\ \midrule
SVM           & 92.5\%   & 92.5\%     & 92.3\%  & 92.3\%    \\
KNN           & 85.0\%   & 86.0\%     & 86.8\%  & 85.8\%    \\
Naive Bayes   & 87.5\%   & 87.6\%     & 87.8\%  & 87.8\%    \\
Decision Tree & 82.9\%   & 83.1\%     & 83.1\%  & 83.0\%    \\
DTW   		  & 78.3\%   & 78.8\%     & 78.6\%  & 78.6\%    \\
DCNN 		  & 91.3\%   & 91.6\%     & 91.6\%  & 91.6\%    \\ \bottomrule

\end{tabular}
}
\label{tab2}
\end{table*}

\begin{figure*}[]
\centering
\subfloat[]{\includegraphics[width=2.3 in]{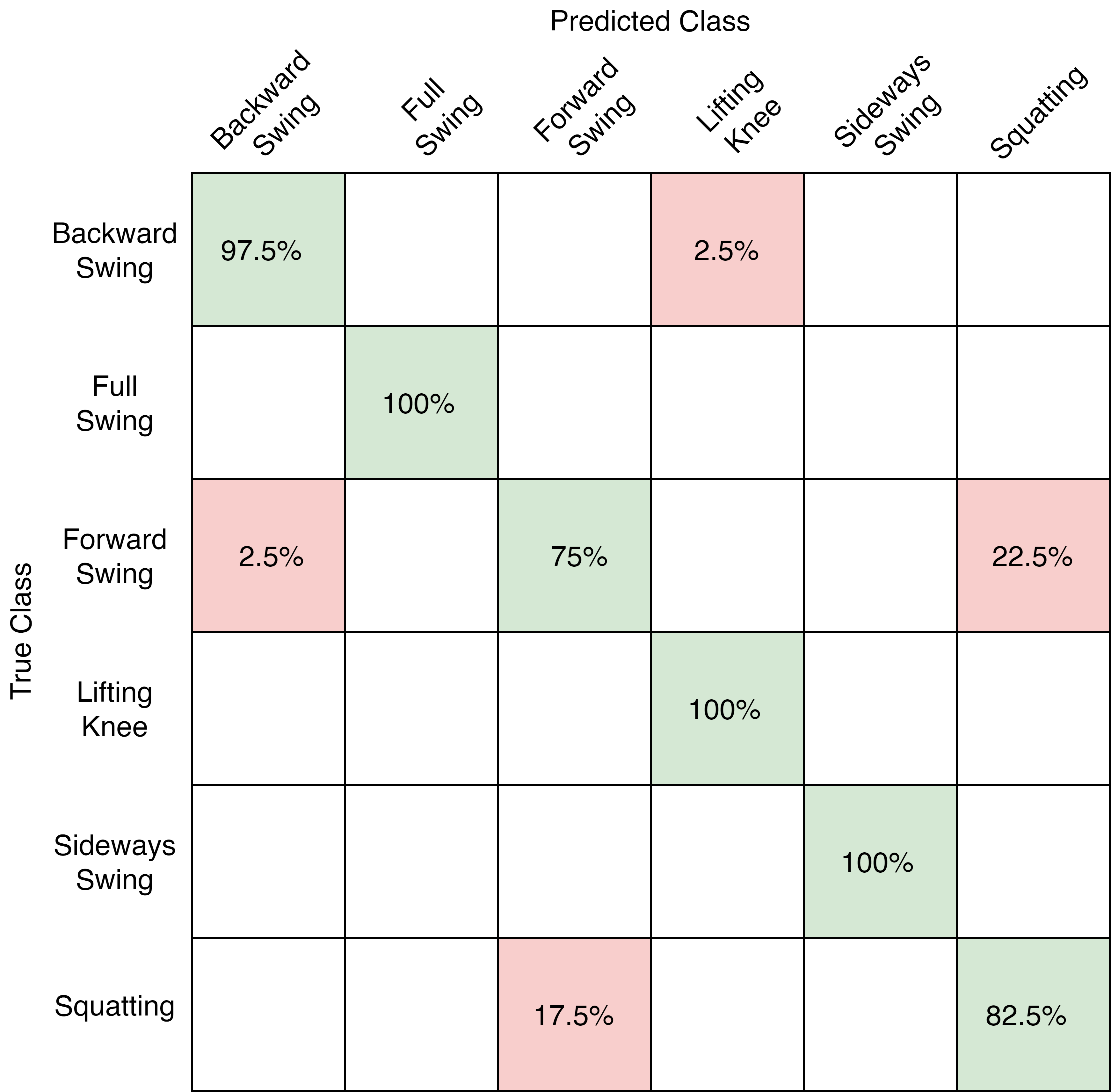}
\label{r1}}\hfill
\subfloat[]{\includegraphics[width=2.3 in]{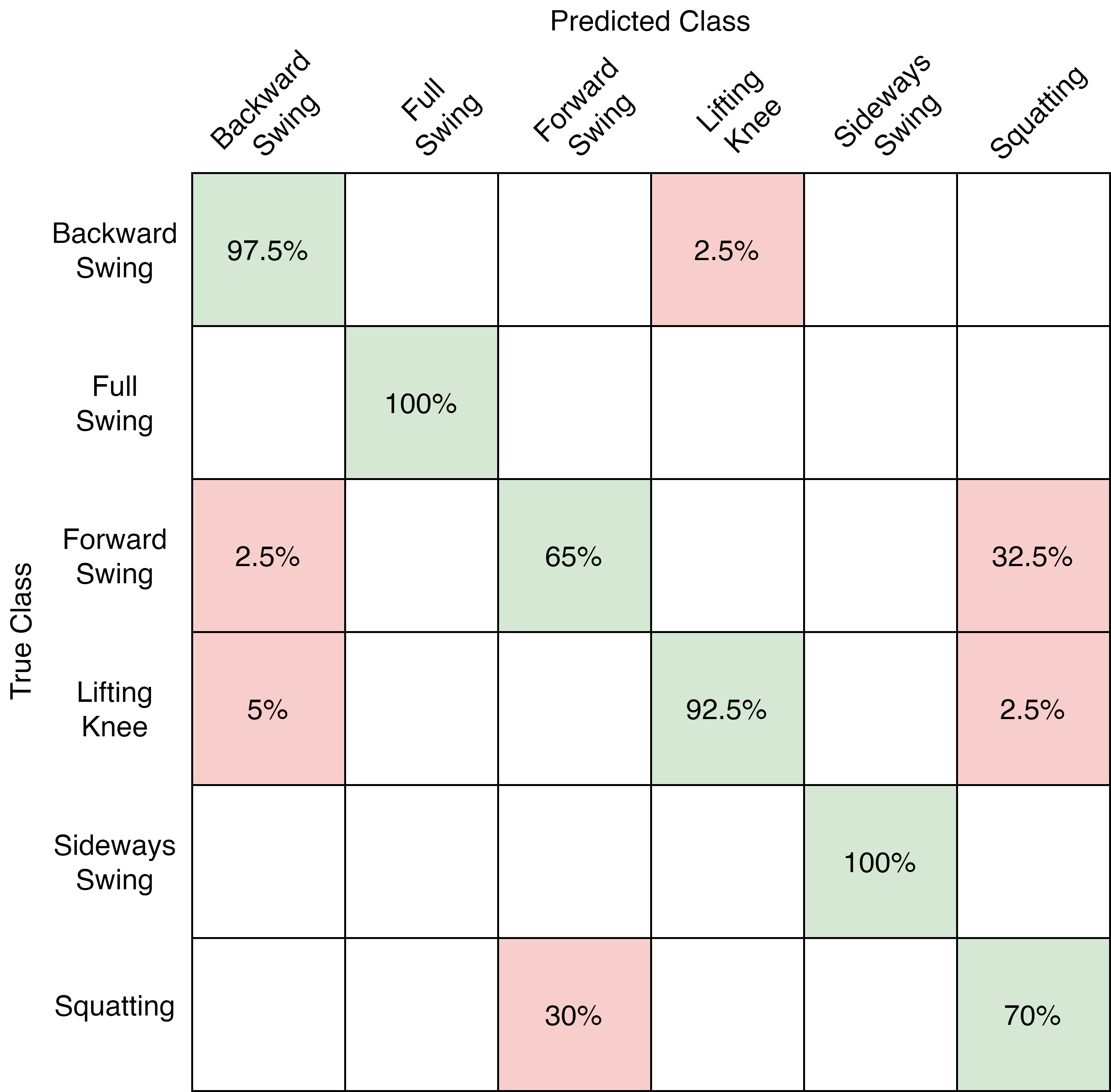}
\label{r2}}\hfill
\subfloat[]{\includegraphics[width=2.3 in]{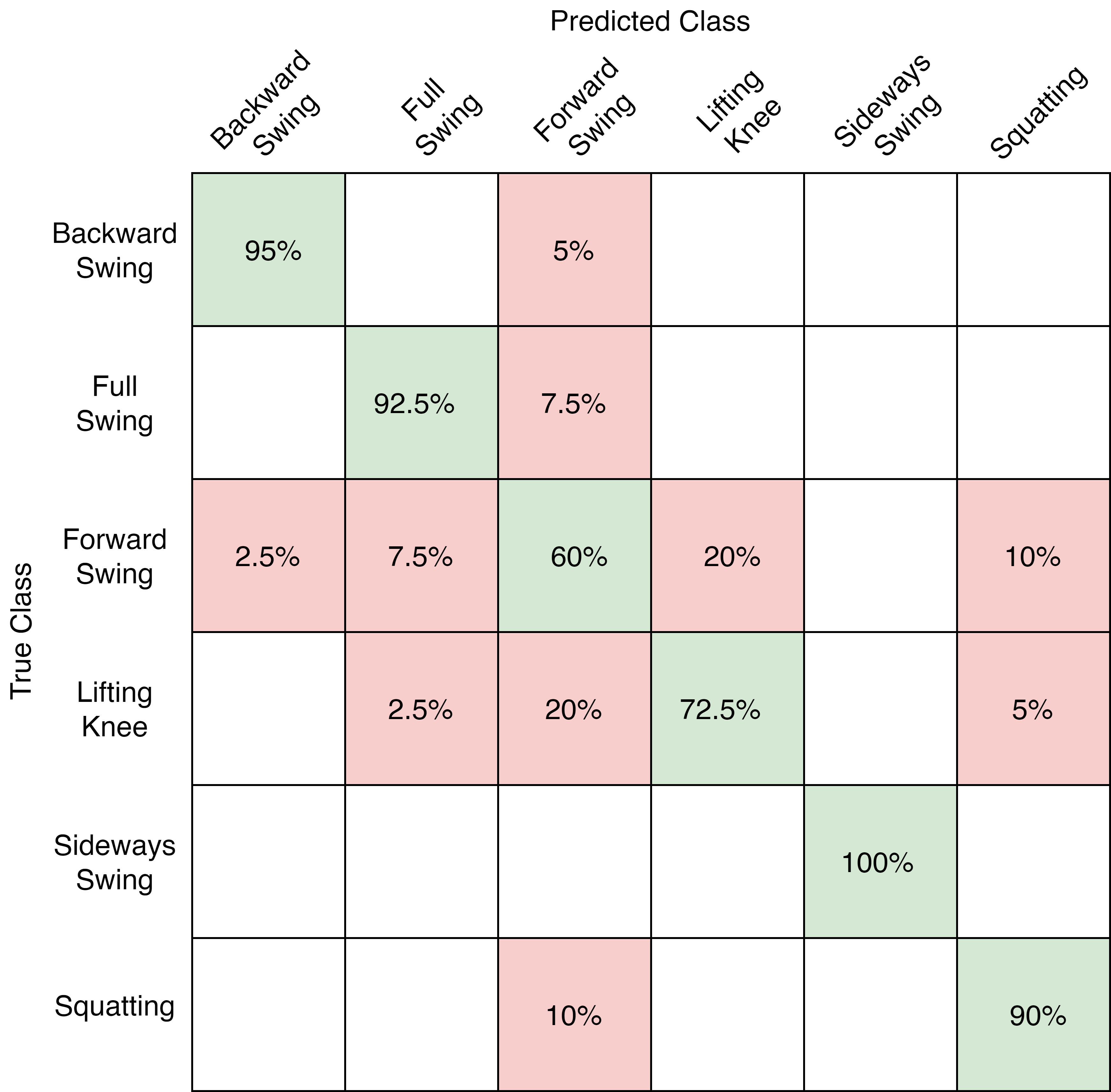}
\label{r3}}\\
\subfloat[]{\includegraphics[width=2.3 in]{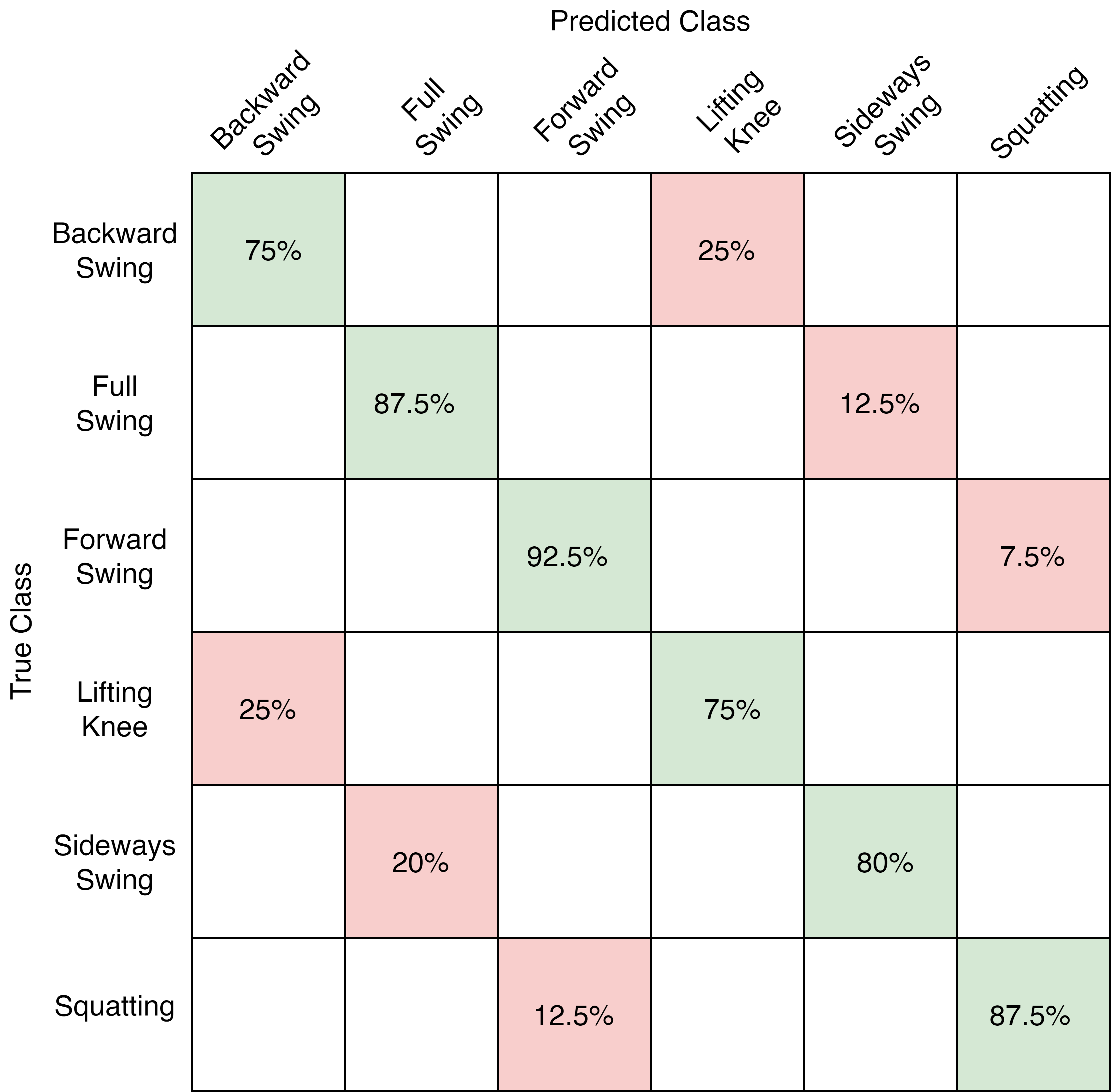}
\label{r4}}\hfill
\subfloat[]{\includegraphics[width=2.3 in]{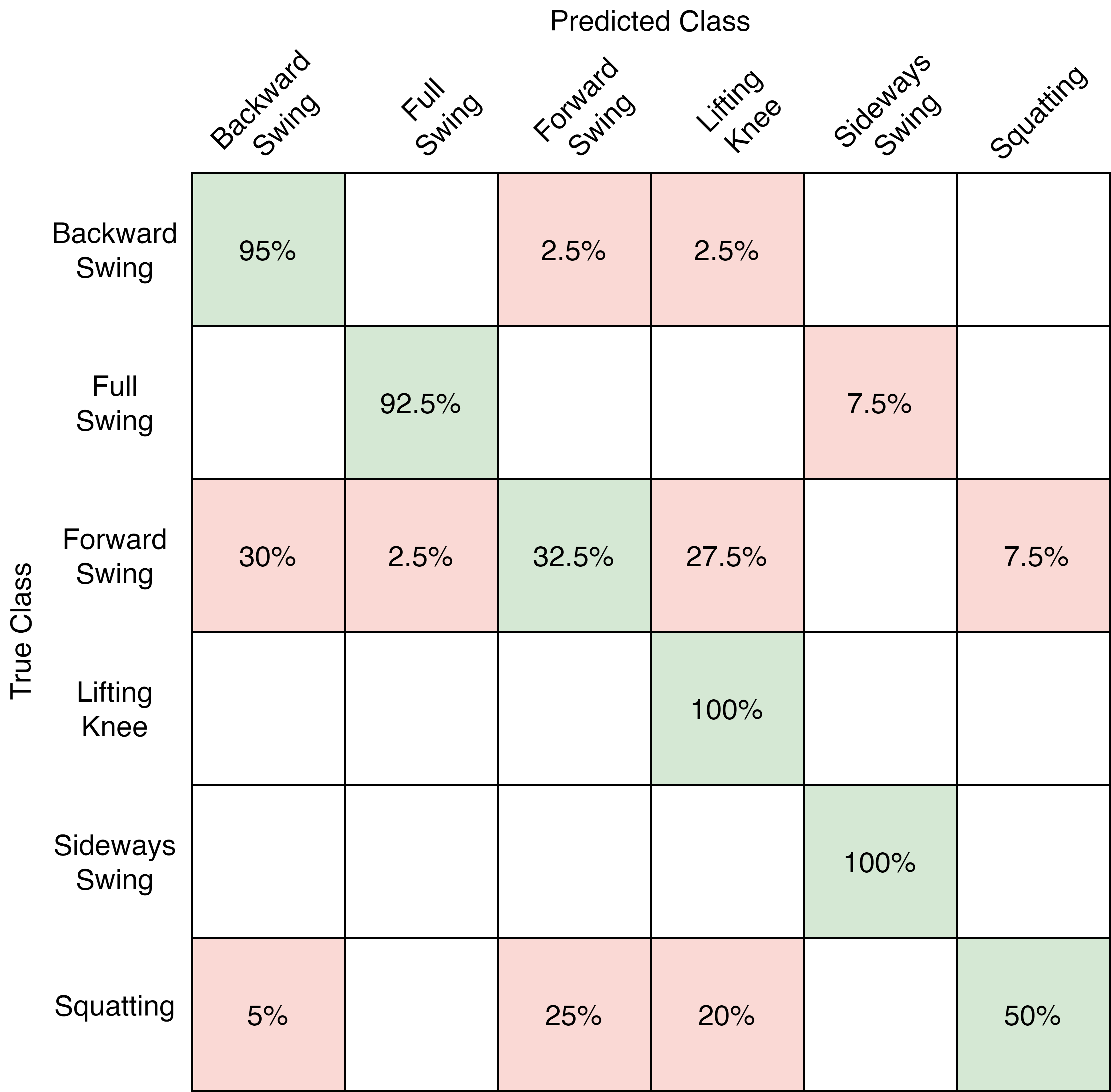}
\label{r5}}\hfill
\subfloat[]{\includegraphics[width=2.3 in]{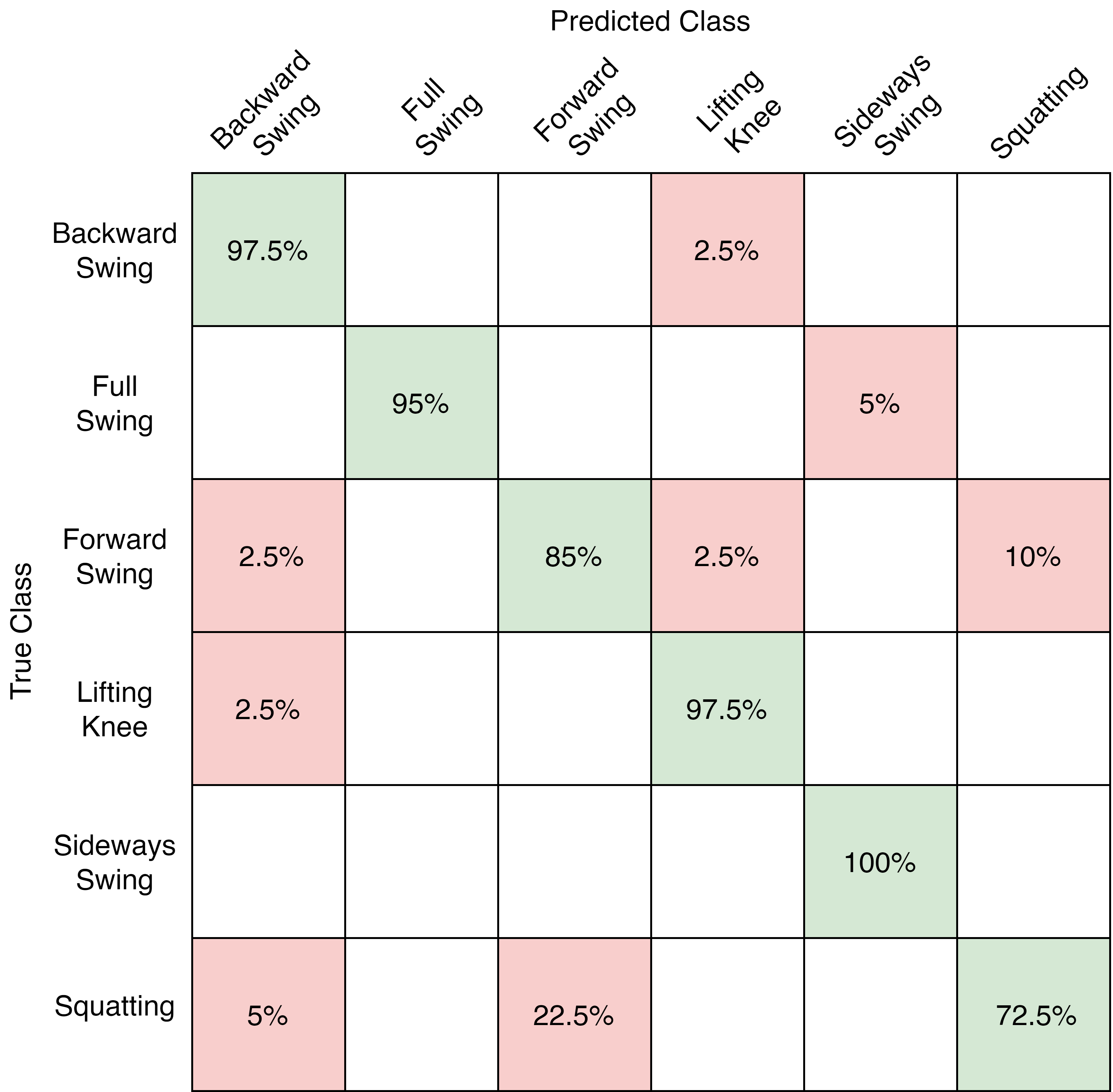}
\label{r6}}
\caption{Confusion matrix for (a) Support Vector Machine (b) Naive Bayes (c) K-Nearest Neighbour (d) Decision Tree \textcolor{black}{(e) DTW  (f) DCNN}}
\label{rr}
\end{figure*}

\begin{figure*}[]
\centering
\subfloat[]{\includegraphics[width=3.5 in]{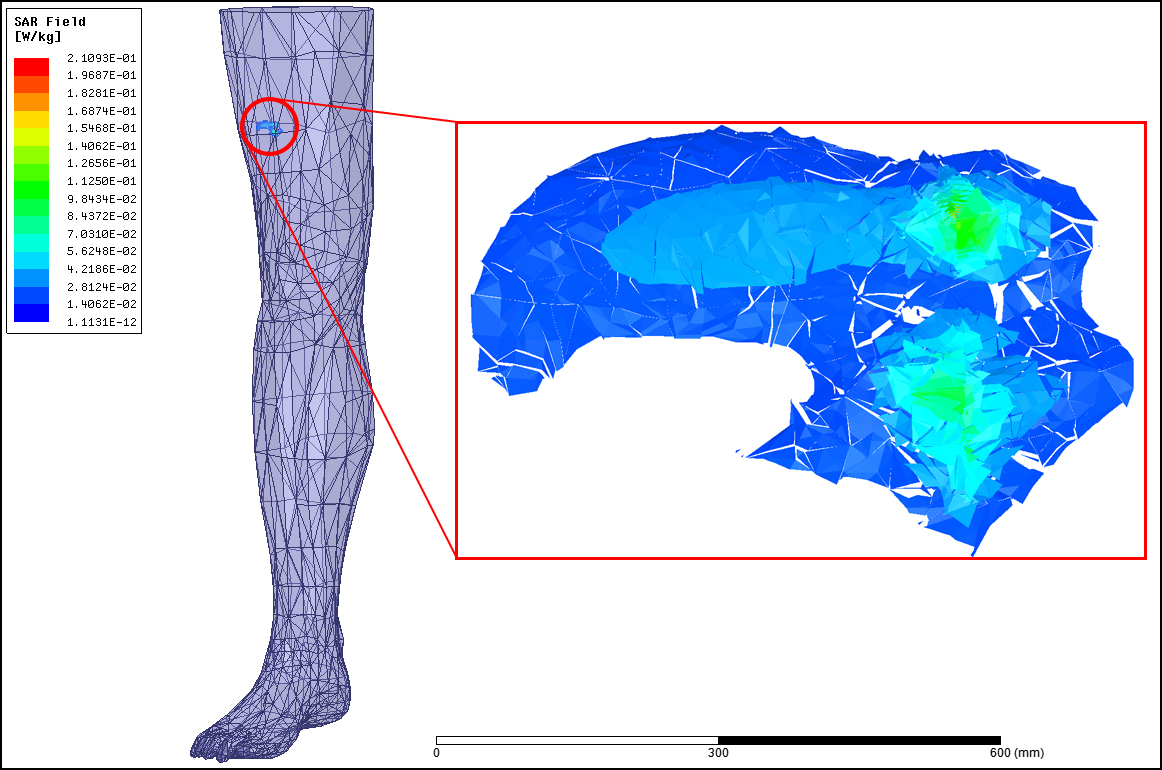}
\label{sar1}}
\caption{Specific absorption rate at 2.4 GHz on the right thigh}
\label{sar}
\end{figure*}
A hyperplane or group of hyper-planes is created by the support vector machine in a high dimensional space that can be used for classification. A one-vs-one strategy is used for the multi-class classification where the multi-class problem is divided into binary classification problems considering a pair of classes.
The first step in the classification process is to extract the features from the transmission coefficient $S_{21}$ of different activities using DWT. The total number of experimental data-sets recorded is 720 (6 subjects $\times$ 6 activities $\times$ 10 trials $\times$ 2 samples). The values in the data-set are normalized from 0 to 1. The feature data-set of different activities are then trained using different classifiers. \textcolor{black}{In our study, we have total 120 signals for each activity and 6 features, the feature set for each activity is of the size 120$\times$6. We take mean of the feature set along the column to get the mean of the features for each activity.} Table \ref{tab1} shows the mean value of the time domain features for the six activities of the DWT coefficients when 4 sec time window of the signal is used. We have used the classification learner toolbox in MATLAB for training. A 5-fold cross-validation method has been used to measure the performance of the model which means that 20\% of the data is used for testing.

To assess the performance in classifying the different leg-based activities, we have employed several commonly used classifiers including SVM, naive bayes, K-nearest neighbours (KNN) and decision trees along with DWT. Classification using dynamic time warping (DTW) \cite{p2_a} and deep convolutional neural network (DCNN) \cite{p2_b} is also implemented based on $S_{21}$ signal. \textcolor{black}{In DTW, the measured $S_{21}$ signals are compared with the reference signals which are already recorded for the activities. The architecture of the DCNN used consists of four convolutional layers with pooling layers after successive convolution layers, followed by two fully connected layers. The training parameters for the above mentioned classifiers are given in Table \ref{para}.} The study has been empirically evaluated by performing experiments with the creeping wave propagation.

\section{Results and Discussion}
%\vspace{-0.5mm}
\textcolor{black}{Table \ref{tab2} shows the performance comparison of the 4 classifiers: SVM, KNN, Naive Bayes and decision tree applied along with discrete wavelet transform, and DTW and DCNN in classifying the six activities based on the magnitude of the creeping wave around the thigh for the six subjects when 4 sec time window of the signal is used. It is observed from the analysis that SVM along with DWT outperforms in classifying different leg-based activities in almost every cases.} 

A confusion matrix is provided in Fig. \ref{rr} to further investigate the performance of the classifiers. The confusion matrix presents the predicted activity (top row) and the true activity (left column) and summarizes the prediction results of our classification problem. For the SVM, it is seen from Fig. \ref{r1} that forward swing is misclassified as backward swing for 2\% and squatting for 22.5\%. Similarly, backward swing is misclassified as lifting knee for 2.5\%, and squatting is misclassified as forward swing for 17.5\%. Out of these, squatting and forward swing are the most confusing activities. Performance parameters including precision, recall, and F-measure for each of the classifiers are calculated and shown in Table \ref{tab2}.

\begin{figure}[]
    \centering
    \includegraphics[width=3.5 in]{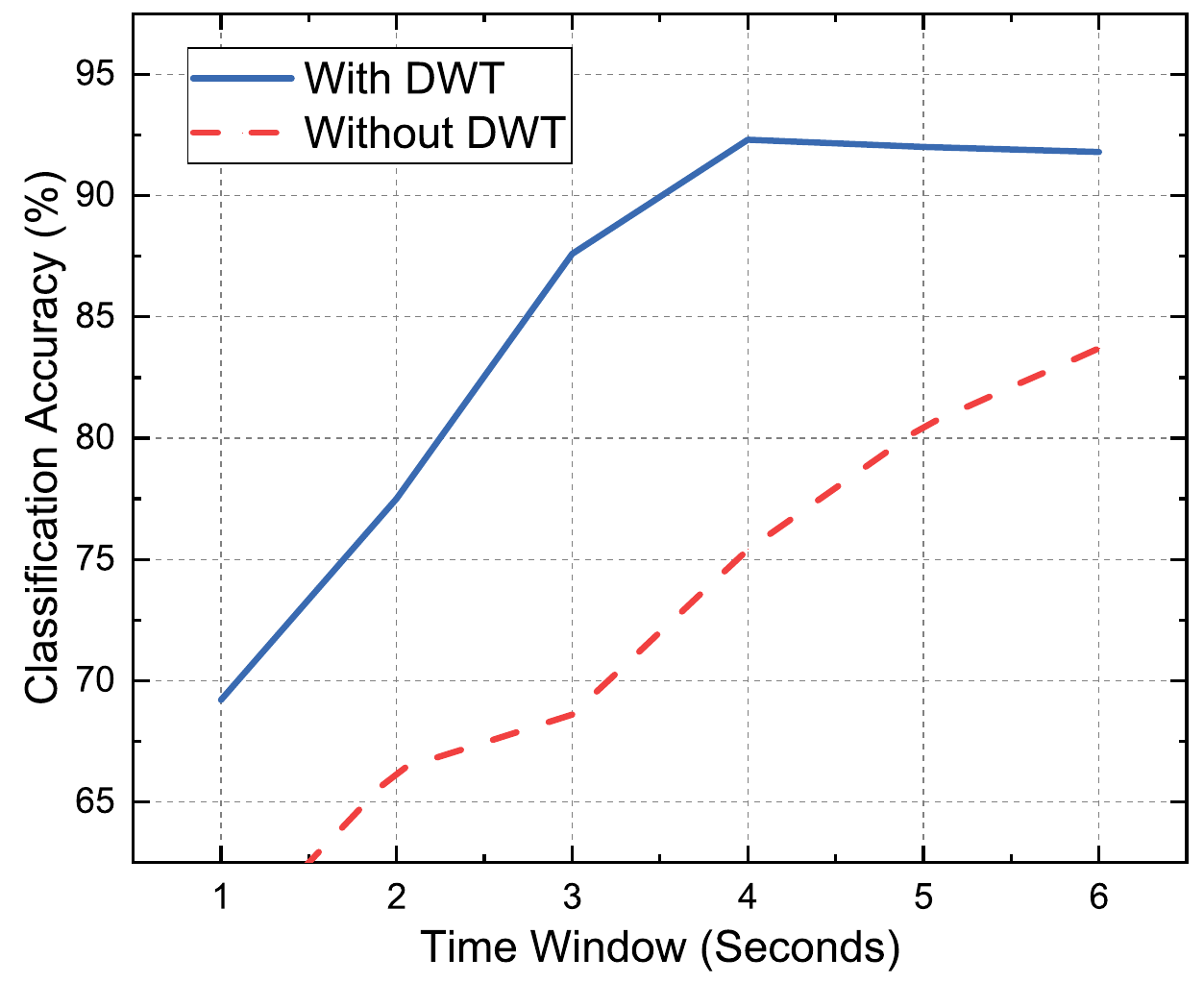}
    \caption{\textcolor{black}{Six leg-based activity classification accuracy with time window with support vector machine}}
    \label{acc}
\end{figure}

Furthermore, the effect of the duration of the activity on the accuracy of the classification is explored. \textcolor{black}{Fig. \ref{acc} shows the classification accuracy vs time duration of the activity. It is preferred to have the time window sufficiently large so as to capture one period of the activity. In our case, the activity period ranges from 2 to 3 seconds. The aim was to find out the minimum time window that would give the highest accuracy. The graph indicates that the classification accuracy of 92.5\% is achieved using DWT for a time window of 4 seconds, whereas 83.7\% is achieved without DWT for a time window of 6 seconds.} \textcolor{black}{For both the cases, the same set of features is used as listed in Table \ref{tab1}.} 

The specific absorption rate analysis of the PET antenna is also performed using a human leg model at a distance of 0.5 mm from the antenna with an input power of 100 mW. The peak SAR value reported is 0.211 W/kg as shown in Fig. \ref{sar}, which is within the safe limits prescribed by International Electrotechnical Commission (IEC) or Federal Communications Commission (FCC) standards.

\section{Conclusion}
\textcolor{black}{In this paper}, we have investigated the feasibility of classifying various leg-related activities by using creeping wave propagation. \textcolor{black}{We have found that the on-body creeping wave can be used to distinguish various leg-related activities that would be useful in applications like fitness tracking and rehabilitation monitoring.} Measurements are performed using two PET on-body antennas placed on the thigh, and the results are verified with a theoretical model. The transmission coefficient ($S_{21}$) has been measured using the creeping wave propagation at 2.45 GHz by recording the variation in the channel characteristics as the leg-based activities are performed. To evaluate the classification accuracy of the different activities, four DWT based classifiers and dynamic time warping (DTW) and deep convolutional neural network (DCNN) are implemented.  Among all DWT coefficient based SVM method has offered the highest classification accuracy of 92.5\%. The effect of duration of the activity is also studied and it is observed that a minimum duration of 4 seconds is required to achieve the highest accuracy. Furthermore, the analysis for specific absorption rate is also performed and a peak value of 0.211 W/Kg is obtained.

Future works can include measurements using different types of WBAN antennas with irregular activities over a large number of test subjects. In order to further evaluate the applicability of the proposed method, future work can also include the integration of more operational environments like an open hallway or a room. Furthermore, a cost-effective on-body wearable system can be developed to process the transmission coefficient of the antenna without the need of a VNA.
\bibliographystyle{IEEEtran}
\bibliography{reference}

\vspace*{-10\baselineskip}

\begin{IEEEbiography}[{\includegraphics[width=1in,height=1.25in,clip,keepaspectratio]{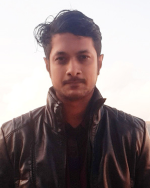}}]{Sagar Dutta}
(S'19) received the B.Engg. degree in electronics and telecommunication engineering from the Gauhati University, Assam, India, in 2015 and the M.Engg. degree in telecommunication engineering from the Asian Institute of Technology, Pathumthani, Thailand, in 2017. He is currently pursuing Ph.D. in electronics and communication engineering from the National Institute of Technology Silchar, Assam, India. His current research interests include wearable antennas, deep learning, digital signal processing and biomedical engineering.

\end{IEEEbiography}

\vspace*{-13\baselineskip}

\begin{IEEEbiography}[{\includegraphics[width=1in,height=1.25in,clip,keepaspectratio]{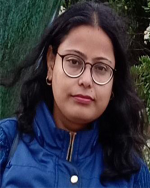}}]{Banani Basu}
(SM'18) received the B.Engg. degree in electronics and communication engineering from the Jalpaiguri Government Engineering College, West Bengal, India, in 2004, the M.Tech. degree in communication engineering from the West Bengal University of Technology, India, in 2008, and the Ph.D. degree in antenna array optimization from the National Institute of Technology Durgapur, West Bengal, India, in 2012.

She is currently an Assistant Professor in the department of electronics and communication engineering with the National Institute of Technology Silchar, Assam, India. Her current research interests include direction-of-arrival estimation of signals and antenna array optimization.

\end{IEEEbiography}

\vspace*{-13\baselineskip}

\begin{IEEEbiography}[{\includegraphics[width=1in,height=1.25in,clip,keepaspectratio]{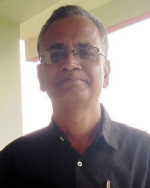}}]{Fazal Ahmed Talukdar}
(M'98) received the B.Engg. degree in electrical engineering from the Regional Engineering College, Silchar, Assam, India, in 1987, the M.Tech. degree in energy studies from the Indian Institute of Technology, Delhi, India, in 1993, and the Ph.D. degree in power electronics from the Jadavpur University, Kolkata, India, in 2003.

He has been with the National Institute of Technology Silchar, Assam, India, since 1991, where he is currently a Professor in the department of electronics and communication engineering. His current research interests include power electronics, signal processing, and analog circuits.

Professor Talukdar is a member of the Project Review Steering Group of the Ministry of Electronics and Information Technology. He is a fellow of the Institution of Engineers (India) and Life Member of Indian Society for Technical Education.
\end{IEEEbiography}
\end{document}